\documentclass[12pt,a4paper]{article}

\usepackage{refmerge}
\usepackage[dvips]{graphicx}
\usepackage{amssymb}

\newcommand{\fourpi}{\ensuremath{\pi^+\pi^-\pi^+\pi^-}}
\newcommand{\eefourpi}{\ensuremath{e^+e^- \rightarrow
    \pi^+\pi^-\pi^+\pi^-}}
\newcommand{\etapp}{\ensuremath{\eta\pi^+\pi^-}}
\newcommand{\fivepi}{\ensuremath{\pi^+\pi^-\pi^+\pi^-\pi^0}}
\newcommand{\re}{\mathop{\mathrm{Re}}}
\newcommand{\im}{\mathop{\mathrm{Im}}}

\title{Observation of the $\phi \to \fourpi$ Decay}
\author{
R.R.~Akhmetshin\footnote{Budker
Institute of Nuclear Physics, Novosibirsk, 630090, Russia},
E.V.~Anashkin\footnotemark[1],
M.~Arpagaus\footnotemark[1], \and
V.M.~Aulchenko\footnotemark[1]
\footnote{Novosibirsk State University, Novosibirsk, 630090, Russia},
V.Sh.~Banzarov\footnotemark[1], 
L.M.~Barkov\footnotemark[1] \footnotemark[2], \and
N.S.~Bashtovoy\footnotemark[1],
A.E.~Bondar\footnotemark[1] \footnotemark[2],
D.V.~Bondarev\footnotemark[1], \and
A.V.~Bragin\footnotemark[1],  
D.V.~Chernyak\footnotemark[1],
S.I.~Eidelman\footnotemark[1] \footnotemark[2], \and
G.V.~Fedotovitch\footnotemark[1] \footnotemark[2],  
N.I.~Gabyshev\footnotemark[1], 
A.A.~Grebeniuk\footnotemark[1], \and
D.N.~Grigoriev\footnotemark[1],
V.W.Hughes\footnote{Yale University, New Haven, CT 06511, USA},
F.V.Ignatov\footnotemark[1] \footnotemark[2],
P.M.Ivanov\footnotemark[1], \and
S.V.~Karpov\footnotemark[1],
V.F.~Kazanin\footnotemark[1] \footnotemark[2],
B.I.~Khazin\footnotemark[1] \footnotemark[2],
I.A.~Koop\footnotemark[1], \and
M.S.~Korostelev\footnotemark[1],
P.P.~Krokovny\footnotemark[1] \footnotemark[2],
L.M.~Kurdadze\footnotemark[1] \footnotemark[2], \and
A.S.~Kuzmin\footnotemark[1] \footnotemark[2],  
I.B.~Logashenko\footnotemark[1],
P.A.~Lukin\footnotemark[1], \and
K.Yu.~Mikhailov\footnotemark[1] \footnotemark[2],
A.I.~Milstein\footnotemark[1] \footnotemark[2],
I.N.~Nesterenko\footnotemark[1], \and
V.S.~Okhapkin\footnotemark[1],
A.V.~Otboev\footnotemark[1],
E.A.Perevedentsev\footnotemark[1] \footnotemark[2], \and
A.S.~Popov\footnotemark[1] \footnotemark[2],
T.A.~Purlatz\footnotemark[1] \footnotemark[2], 
S.I.Redin\footnotemark[1],
N.I.~Root\footnotemark[1] \footnotemark[2], \and
A.A.~Ruban\footnotemark[1],
N.M.~Ryskulov\footnotemark[1],
A.G.~Shamov\footnotemark[1],  \and
Yu.M.~Shatunov\footnotemark[1],
B.A.~Shwartz\footnotemark[1] \footnotemark[2],
A.L.~Sibidanov\footnotemark[1] \footnotemark[2], \and
V.A.~Sidorov\footnotemark[1], 
A.N.~Skrinsky\footnotemark[1], 
V.P.~Smakhtin\footnotemark[1],\and
I.G.~Snopkov\footnotemark[1], 
E.P.~Solodov\footnotemark[1] \footnotemark[2],
P.Yu.~Stepanov\footnotemark[1], \and
A.I.~Sukhanov\footnotemark[1],
J.A.Thompson\footnote{University of Pittsburgh, Pittsburgh, PA 15260, USA},
V.M.~Titov\footnotemark[1], \and
A.A.~Valishev\footnotemark[1],
Yu.V.~Yudin\footnotemark[1],
S.G.~Zverev\footnotemark[1]
}

\date{\today}

\begin{document}

\maketitle

\newpage

\begin{abstract}
Using 11.6 $\mbox{pb}^{-1}$ of data collected in the energy
range 0.984--1.06 GeV by CMD-2 at VEPP-2M, the cross section of the
reaction \eefourpi\ has been studied. For the first time an interference
pattern was observed in the energy dependence of the cross section 
near the $\phi$ meson. The branching ratio of the $\phi\to\fourpi$ decay
double suppressed by the
G-parity and OZI rule is measured

\begin{displaymath}
  Br(\phi\to\fourpi) = (3.93 \pm 1.74 \pm 2.14) \cdot 10^{-6}.
\end{displaymath}
The upper limits have been placed for the decays $\phi\to\fivepi$ and
$\phi\to\etapp$
\begin{eqnarray}
  Br(\phi\to\fivepi) & < & 4.6 \cdot 10^{-6}\quad 90\%\,\mbox{CL},
  \nonumber \\
  Br(\phi\to\etapp) & < & 1.8 \cdot 10^{-5}\quad 90\%\,\mbox{CL}.
  \nonumber
\end{eqnarray}
\end{abstract}

%===================================================================

\section{Introduction}

Production of four pions in $e^+e^-$ annihilation is now well studied
in the c.m. energy range 1.05 to 2.5 GeV (see \cite{our} and references
therein).
Results on the measurements of the cross section of the
reaction \eefourpi\ in the energy range from 0.60 to 0.97 GeV
as well as the probability of the $\rho^0$ meson decay into the \fourpi\ 
final state were recently presented by the CMD-2 group \cite{rho4pi}.
However, the behavior of the cross section of the process \eefourpi\
in the vicinity of the $\phi$ meson has not been as well studied.
In the previous experiments performed in Orsay \cite{m2n,dm1} and in
Novosibirsk \cite{olya,nd,cmd} the cross section was measured at
single points at $E_{cm} \approx m_{\phi}$. Because of the small data samples
in these experiments, no detailed studies of the cross section structure
in the $\phi$ meson region could be made. Under the assumption that the
visible cross section is due to the $\phi$ decay, an upper limit
was set on the value of the decay probability $\phi \to \fourpi$
\cite{dm1}. The intensity of this decay is of interest since 
it is twice suppressed, by G-parity and the OZI rule.  

Two other rare $\phi$ decays which violate the OZI rule are 
$\phi\to\etapp$ (also forbidden by G-parity) and $\phi\to\fivepi$.
A search for the decay $\phi\to\etapp$ based on part of the
total data sample was performed by CMD-2 \cite{3pi} using the $\eta \to \gamma
\gamma$ decay mode. No events of this decay were observed and an
upper limit was placed.
Earlier the CMD group set an upper limit for the branching ratio
of the decay $\phi\to\fivepi$ \cite{cmd}.

In 1992 the upgraded high luminosity collider VEPP-2M resumed its
operation at the Budker Institute of Nuclear Physics in Novosibirsk
\cite{vepp}. Two modern detectors CMD-2 \cite{cmddec} and SND
\cite{SND} started a series of experiments which include various high
precision measurements in the c.m. energy range from the threshold of
hadron production to 1.4 GeV. High data samples collected by both detectors 
in the $\phi$ meson energy range allowed the first observation of 
various rare decay modes among which are G-parity and OZI rule suppressed
decays to $\pi^+\pi^-$ \cite{pipicmd,pipisnd} and $\omega \pi^0$ \cite{ompi}. 

In this paper we extend the analysis of the process \eefourpi\ 
started by CMD-2 in Refs. \cite{our,rho4pi} to the $\phi$ meson 
c.m. energy range from 0.984 to 1.06 GeV. The high integrated
luminosity allowed the observation of 
the clear interference pattern at $E_{cm} \approx m_{\phi}$ 
indicating the presence of the decay $\phi \to \fourpi$.
The same data sample was used to search for the decays $\phi \to
\etapp$ and $\phi \to \fivepi$. 

%===================================================================

\section{Experiment and data analysis}\label{sec:analysis}

Three scans of the energy range from 0.984
to 1.06 GeV were performed in winter 1997--1998. The scan step
was 1 MeV near the $\phi$ meson
(1.016--1.023 GeV) and 6--10 MeV outside the resonance.
Some luminosity has also been collected at 1.019 and
1.020 GeV before the main scans and at 1.017 and 1.020 GeV after them.
For the final analysis data samples from the same energy points 
of different scans were combined.

The general purpose detector CMD-2 has been described in detail 
elsewhere \cite{cmddec}. It consists of a drift chamber (DC) 
\cite{dcdec} and a 
proportional Z-chamber \cite{zcdec}, both used for the trigger, 
and both inside 
a thin (0.4 $X_0$) superconducting solenoid with a field of 1 T.

The barrel calorimeter \cite{csidec} which is placed outside the solenoid,
consists of 892 CsI crystals of $6\times 6\times 15$ cm$^3$ size and covers
polar angles from $46^\circ$ to $134^\circ$. The energy resolution for
photons is about 9\% in the energy range from 50 to 600 MeV. The angular
resolution is about 0.02 radians.

The end-cap calorimeter \cite{bgodec} which is placed inside the solenoid,
consists of 680 BGO crystals of $2.5\times 2.5\times 15$ cm$^3$ size and covers
forward-backward polar angles from 16$^\circ$ to 49$^\circ$ and
from 131$^\circ$ to 164$^\circ$. The energy and angular resolution
varies from 8\% to 4\% and from 0.03 to 0.02 radians respectively
for the photons in the energy range from 100 to 700 MeV.

The luminosity was determined from the events of Bhabha scattering
at large angles \cite{prep99}.

The collider energy was roughly set $(\delta E/E
\lesssim 10^{-3})$ by the dipole magnet currents. 
In the energy range 1.010 to 1.028 GeV of the main scans the beam energy 
was more precisely $(\delta E/E \lesssim 10^{-4})$ determined by measuring
the average momentum $p_{av}$ of $K^+K^-$ pairs in the DC:
$E^{K^+K^-} = \sqrt{p^2_{av} + M^2_{K}} + \Delta$. Here $\Delta$
is a correction for the contributions of kaon ionization losses inside
the detector and radiative losses of initial electrons. Its magnitude
depends on $p_{av}$ and varies from 5 to 3 MeV for $p_{av}$ in the
range 80 to 130 MeV/c (see \cite{prep99} for more detail).
Then the effective beam energy at each energy point
was determined by averaging $E^{K^+K^-}$ weighted by
the integrated luminosity:
\begin{displaymath}
  E_i= \sum_{j=1}^{3}L_{ij}E^{K^+K^-}_{ij}\bigg/\sum_{j=1}^{3}L_{ij}\,,
\end{displaymath}
where $E^{K^+K^-}_{ij}$ and $L_{ij}$ are the kaon energy and the
integrated luminosity measured at the $i^{th}$ energy point of the
$j^{th}$ scan.

The first two columns of Table~\ref{tab:xsec} present the
corresponding energy values
and integrated luminosities. The total integrated
luminosity was $11.63 \,\mbox{pb}^{-1}$.

The analysis of the reaction \eefourpi\ was performed similarly to our
analysis described in \cite{rho4pi}. However, because of the completely
different background situation, other methods of background
suppression were used. Events with four charged tracks 
coming from the interaction region were selected:
\begin{itemize}
\item
the impact parameter
of each track $r_{min}$ is less than 1 cm
\item
the vertex coordinate
along the beam axis $z_{vert}$ is within $\pm 10$ cm.
\end{itemize}
To have good reconstruction efficiency, tracks were also required to
cross at least two superlayers of the drift chamber: $|\cos\theta| < 0.8$.

For selected events a kinematic fit was performed,
assuming that all tracks are pions
and under the constraint that the sum  of the 3-momenta
$\sum_{i=1}^{4}{\vec{p}_i}=0$.
Then the requirement that the fit quality $\chi^{2}_{4\pi}$ / n.d.f. $<$ 
100/3  was applied. 
%which characterizes the quality of the kinematic fit. 
This condition has high efficiency (about 95\%) for the process
under study and rejects about 70\% of the background reactions:
\begin{eqnarray}
\phi & \to & K_S^{0}K_L^{0} \label{eq:kskl}
\end{eqnarray}
and
\begin{eqnarray}
\phi &\to& K^+K^- \,. \label{eq:kpkm}
\end{eqnarray}
Further analysis was performed using the normalized
``apparent energy'':
\begin{eqnarray}
  \varepsilon_{app}&=&\frac
  {\sum_{i=1}^{4}{\sqrt{\vec{p}_i^2+m_{\pi}^2}}}
  {2E_{beam}}. \nonumber
\end{eqnarray}
\begin{figure}
\begin{center}
  \includegraphics[height=0.6\textwidth]{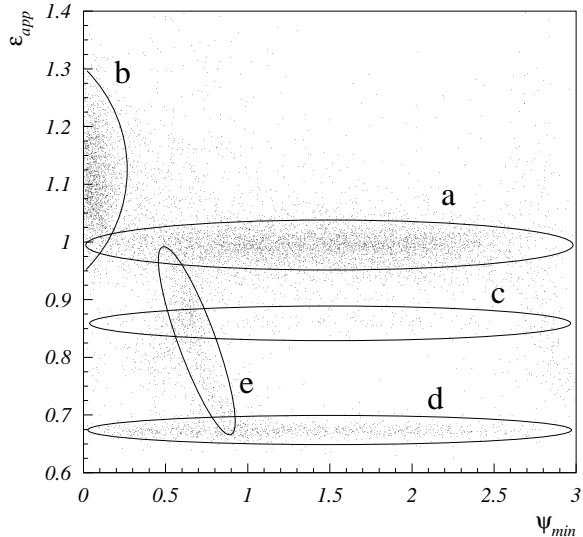}
     \caption{Distribution of the normalized
      apparent energy versus the minimum space angle between tracks with
      opposite charges}
    \label{fig:eapp-psi}
\end{center}
\end{figure}
Figure \ref{fig:eapp-psi} shows the distribution of
$\varepsilon_{app}$ versus the minimum space angle between the tracks
with the opposite charges $\psi_{min}$. A band with $\varepsilon_{app}
\approx 1$ corresponding to \fourpi\ events is clearly observed in the
region ``a''. The lower part of the region ``b'' is populated by
events from the process:
\begin{eqnarray}
  e^+e^- &\to& \pi^+\pi^-\pi^0\,,\pi^0 \to e^+e^-\gamma\,, \label{eq:3pi}
\end{eqnarray}
while events from the processes:
\begin{eqnarray}
  e^+e^- &\to& e^+e^-\gamma\,, \label{eq:eeg} \\
  e^+e^- &\to& \pi^+\pi^-\gamma \label{eq:ppg}
\end{eqnarray}
with the subsequent photon conversion into an $e^+e^-$-pair at
the beam pipe, fall into the upper part of the region ``b''.
Events of the $\phi$ meson decay (\ref{eq:kskl}),
where $K_S^0$ and $K_L^0$ decay to $\pi^+\pi^-$ and $\pi^+\pi^-\pi^0$
respectively, contribute to the region ``c''.
Events of  another $\phi$ meson decay (\ref{eq:kpkm}),
where  products of kaon nuclear interactions scatter
back to the drift chamber and induce two ``extra'' tracks, fall in
the region ``e''. Events with the decay of one of the kaons $K^{\pm} \to
\pi^{\pm}\pi^+\pi^-$ populate the region ``d''.
Thus, using the parameter $\varepsilon_{app}$, we are able to separate
events of the classes ``c'' and ``d'' from \fourpi\ events, whereas
events from the regions ``b'' and ``e'' constitute the background to
the process under study.

\begin{figure}
\begin{center}
  \includegraphics[height=0.49\textwidth]{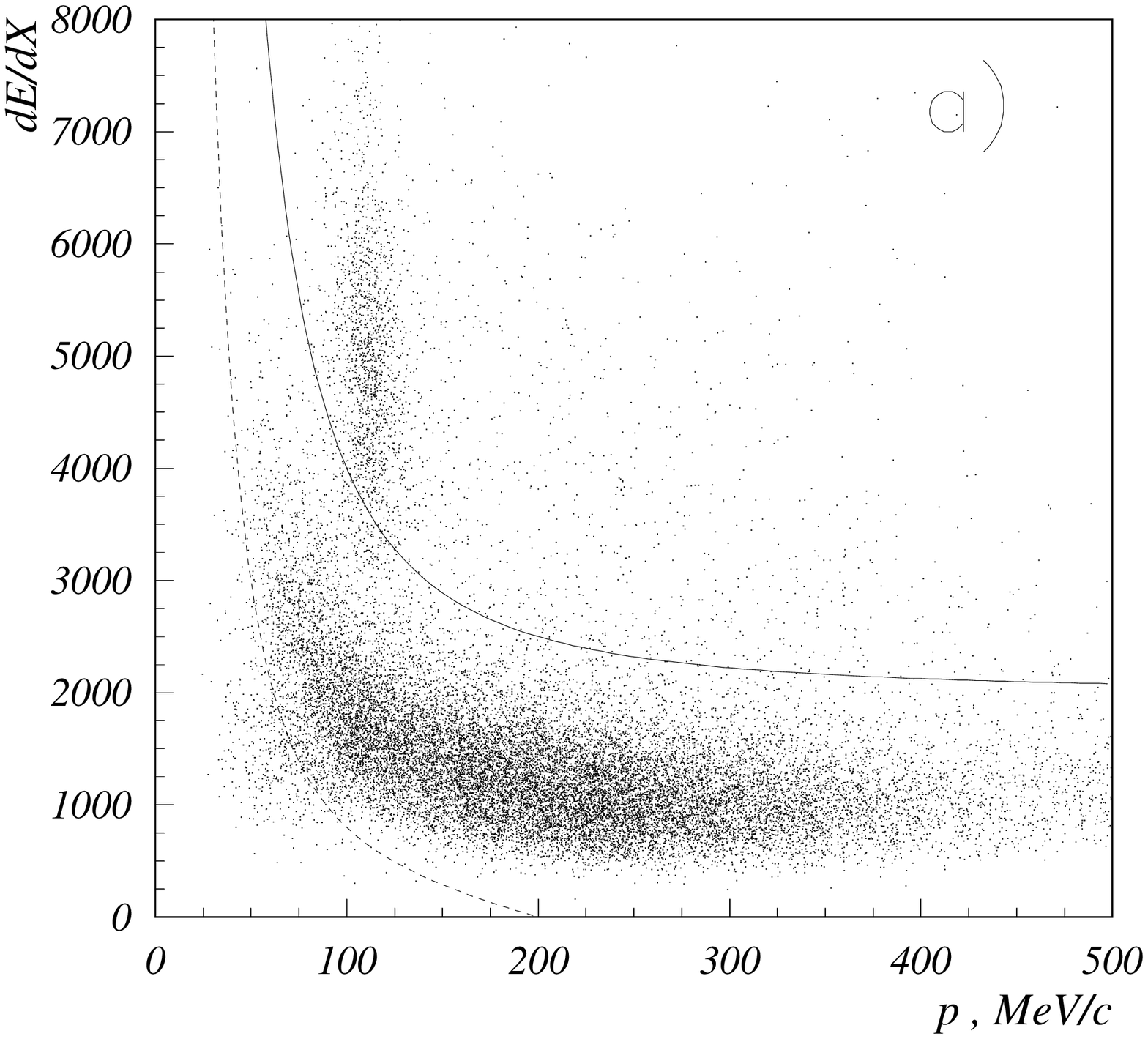}
  \hfill
  \includegraphics[height=0.49\textwidth]{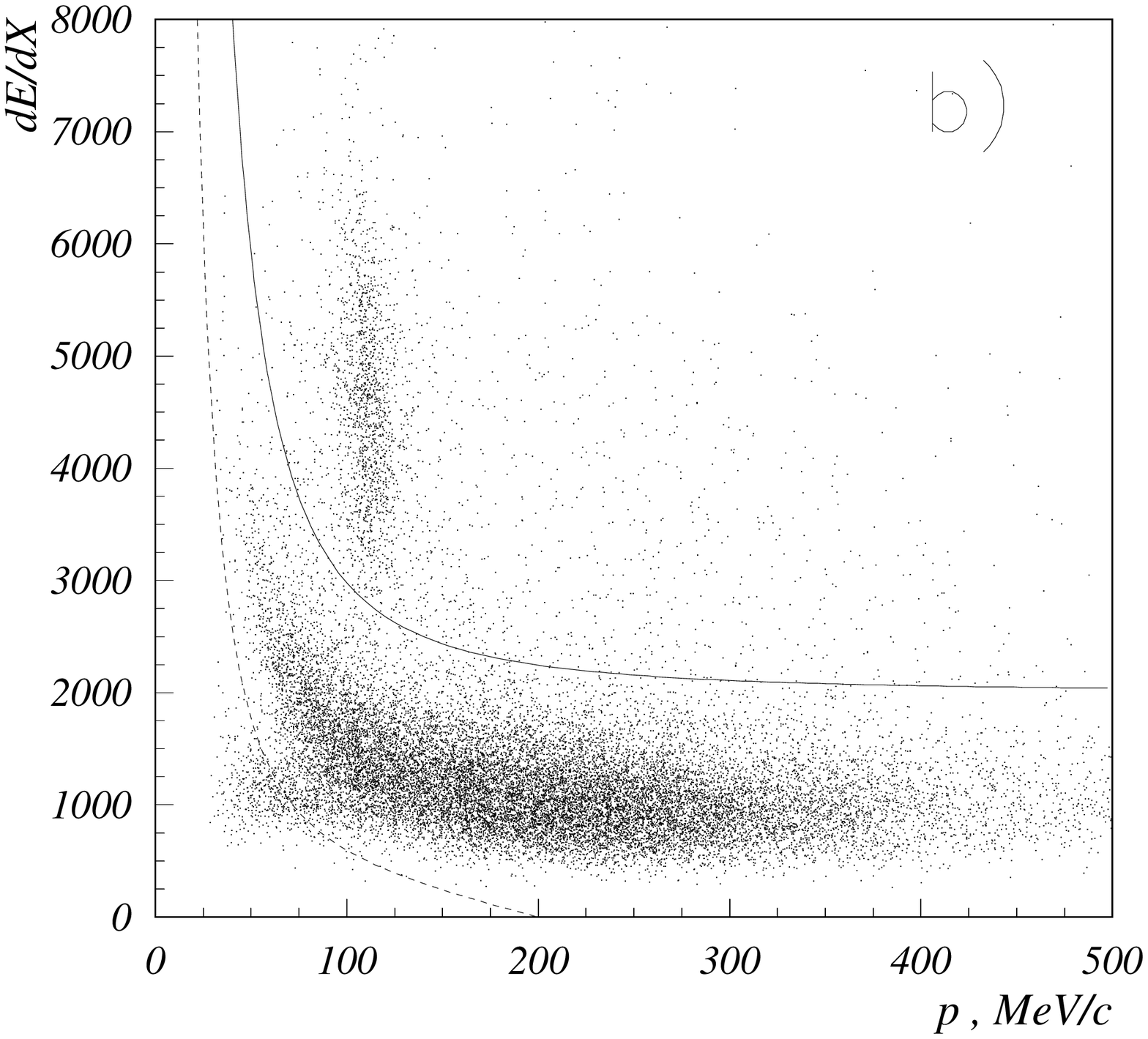}
  \\
     \caption{Ionization losses (arbitrary units) versus the
      track momentum: a) positively charged tracks, b) negatively charged
      tracks. The solid and dashed lines show the selection boundaries}
     \label{fig:dedx}
\end{center}
\end{figure}

To suppress the background from the process $\phi \to K^+K^-$ with the
nuclear interaction of kaons, ionization losses $dE/dx$ of
the tracks measured in the drift chamber \cite{dcdec} were used.
Figure \ref{fig:dedx} shows the scatter plot of the ionization losses
$dE/dx$ (in arbitrary units) versus the track momentum $p$ for
positively (a) and negatively (b) charged tracks.
A narrow vertical band with momentum $p \approx 127$ MeV/c and
$dE/dx$ above 3000 in both Figures corresponds to  events with
charged kaons from the process (\ref{eq:kpkm}). 
Thus, selection of events below the solid line 
effectively rejects events from both classes ``d'' and ``e''.
To estimate the number of remaining background events in class ``e'',
events of the process $\phi \to K^+K^-$, $K^{\pm} \to
\pi^{\pm}\pi^+\pi^-$ were selected using the conditions $\varepsilon
<0.7$ and $\psi_{min} > 1$. Then the efficiency
$\varepsilon^{K^{\pm}}_{dE/dx}$ of such a  cut 
was determined for these events. Since in class ``e'' both charged kaons
are detected in the drift chamber, the probability of kaon
misidentification is $(\varepsilon^{K^{\pm}}_{dE/dx})^2$. Using
this probability, the
expected number of remaining background events was found to be
$N_{bg}^{K^+K^-} < 5$.

Events of the reactions (\ref{eq:3pi}), (\ref{eq:eeg}) and
(\ref{eq:ppg}) (region ``b'' in Fig. \ref{fig:eapp-psi}) with 
low momentum of $e^{\pm}$ tracks give a small tail 
in the lower left corner of Fig. \ref{fig:dedx}. 
Selection of events above the dashed line 
suppresses the background from the processes with the photon
conversion into an $e^+e^-$ pair. Additional suppression of the process
(\ref{eq:eeg}) is provided by the requirement $\varepsilon_{4e} <
0.9$, where the parameter $ \varepsilon_{4e} = \sum_{i=1}^{4} |\vec{p}_i| / 2
E_{beam}< 0.9$ is the normalized total energy assuming that all particles
are electrons.

After that, using the conditions $\psi_{min} > 0.3$ and
$|\varepsilon_{app} - 1| < 0.1$, we selected our data sample of about 4200
events consisting mostly of the events of the process \eefourpi.
The number of remaining background events from the process (\ref{eq:3pi}) was
estimated using the $\psi_{min}$ distribution in the region
$\varepsilon_{app} > 1.1$ (see Fig. \ref{fig:eapp-psi}). This area is
populated by events from the reactions (\ref{eq:3pi}), (\ref{eq:eeg}) and
(\ref{eq:ppg}).
Assuming that $\psi_{min}$ distributions  have similar
shape in the regions $\varepsilon_{app} > 1.1$ and $\varepsilon_{app}
< 1.1$, we obtained $N_{bg}^{3\pi} < 80$.

\begin{figure}
\begin{center}
  \includegraphics[height=0.6\textwidth]{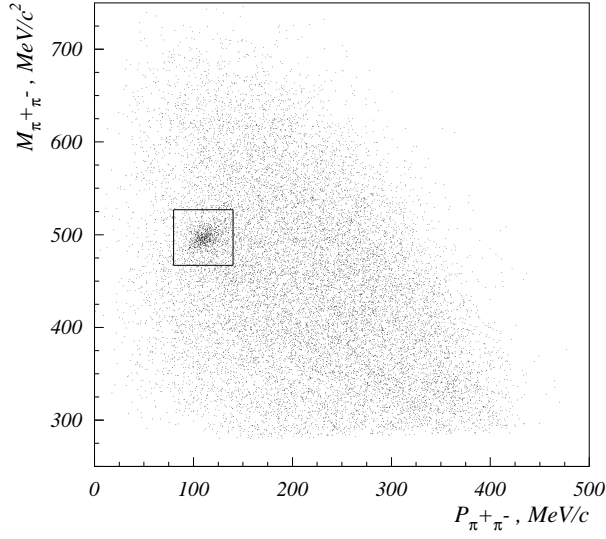}
     \caption{The invariant mass $M_{\pi^+\pi^-}$ of
       pion pairs versus the momentum of the same pair
       $P_{\pi^+\pi^-} = |\vec{p}_{\pi^+} + \vec{p}_{\pi^-} |$}
     \label{fig:mpp-ppp}
\end{center}
\end{figure}

Figure \ref{fig:mpp-ppp} shows the distribution of the invariant
mass $M_{\pi^+\pi^-}$ for pairs of opposite charged pions versus the
total momentum of the same pair 
$P_{\pi^+\pi^-} = |\vec{p}_{\pi^+} + \vec{p}_{\pi^-} |$
for the selected data sample (4 entries per event).
The enhanced concentration of events 
in the region $M_{\pi^+\pi^-} \approx
m_{K_S^0}$ and $P_{\pi^+\pi^-} \approx P_{K_S^0}$ is caused by
the reaction (\ref{eq:kskl}), where $K_S^0$ decays to $\pi^+\pi^-$ and 
$K_L^0$ decays to one of the semileptonic modes
$K_L^0 \to \pi^{\pm}e^{\mp}\nu_{e}$ or
$K_L^0 \to \pi^{\pm}\mu^{\mp}\nu_{\mu}$.
Here $m_{K^{0}_{S}}$=497.67 MeV \cite{pdg} 
and $P_{K^{0}_{S}}=\sqrt{E_{beam}^2-m_{K^{0}_{S}}^2}$ are the 
$K_{S}^0$ meson mass and momentum. 
To reject this type of the background we
excluded events in which at least one of the $\pi^+\pi^-$ pairs satisfies
both of the following conditions:
\begin{eqnarray}
|M_{\pi^+\pi^-}-m_{K_S^0}| &<& 30 \,\mbox{MeV}/c^2, \nonumber \\
|P_{\pi^+\pi^-}-P_{K^0_{S}}| &<& 30 \,\mbox{MeV}/c.
\label{eq:unselks}
\end{eqnarray}
The expected number of remaining background events from this process was
estimated using the complete Monte Carlo simulation (MC) of the CMD-2
detector \cite{cmd2sim} and was found to be $N_{bg}^{K_S^0 K_L^0} < 30$.

%===================================================================

\section{Determination of cross section}

At each energy the cross section of the process
$e^+e^-\rightarrow\pi^+\pi^-\pi^+\pi^-$ was calculated 
using the formula:
\begin{eqnarray}
  \sigma_i & = & \frac{N_{i}}
  {L_i\,\varepsilon_i\,(1+\delta_i)}\,, \label{eq:xsec}
\end{eqnarray}
where $N_{i}$ is the number of selected \fourpi\ events, $L_i$ is the 
integrated luminosity, $\varepsilon_i$ is the detection efficiency, 
and $\delta_i$ is the radiative correction at the $i$-th energy
point.

\begin{figure}
  \begin{center}
  \includegraphics[width=0.6\textwidth]{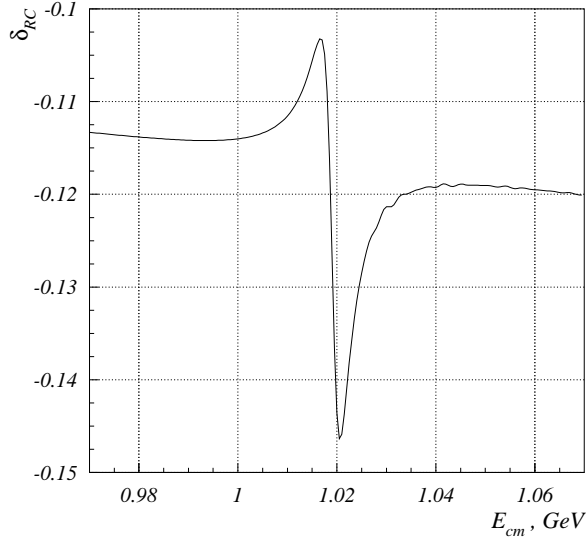}
  \caption{Radiative corrections}\label{fig:rc}
  \end{center}
\end{figure}

The detection efficiency
was determined from MC  assuming the $a_1(1260)\pi$
quasi-twobody production mechanism, which
clearly dominates at higher energy \cite{our}. Comparison
of various experimental distributions with the simulation 
shows that the assumption of the $a_1(1260)\pi$ mechanism
does not contradict the data. 
The detection efficiency decreases with energy, smoothly varying from 
30\% to 27\%.

Radiative corrections were calculated according to \cite{KF}.
Since radiative corrections themselves depend on the energy behavior
of the cross section, the calculation was performed by the iteration
method. ``Visible'' values of the cross section (with $\delta = 0$
in (\ref{eq:xsec})) were used as the first approximation. Then the
cross section was recalculated with the new values of the radiative
corrections and the whole procedure was repeated until the convergence
was reached.
Figure \ref{fig:rc} demonstrates the energy dependence of radiative
corrections.

\begin{figure}
\begin{center}
  \includegraphics[height=0.6\textwidth]{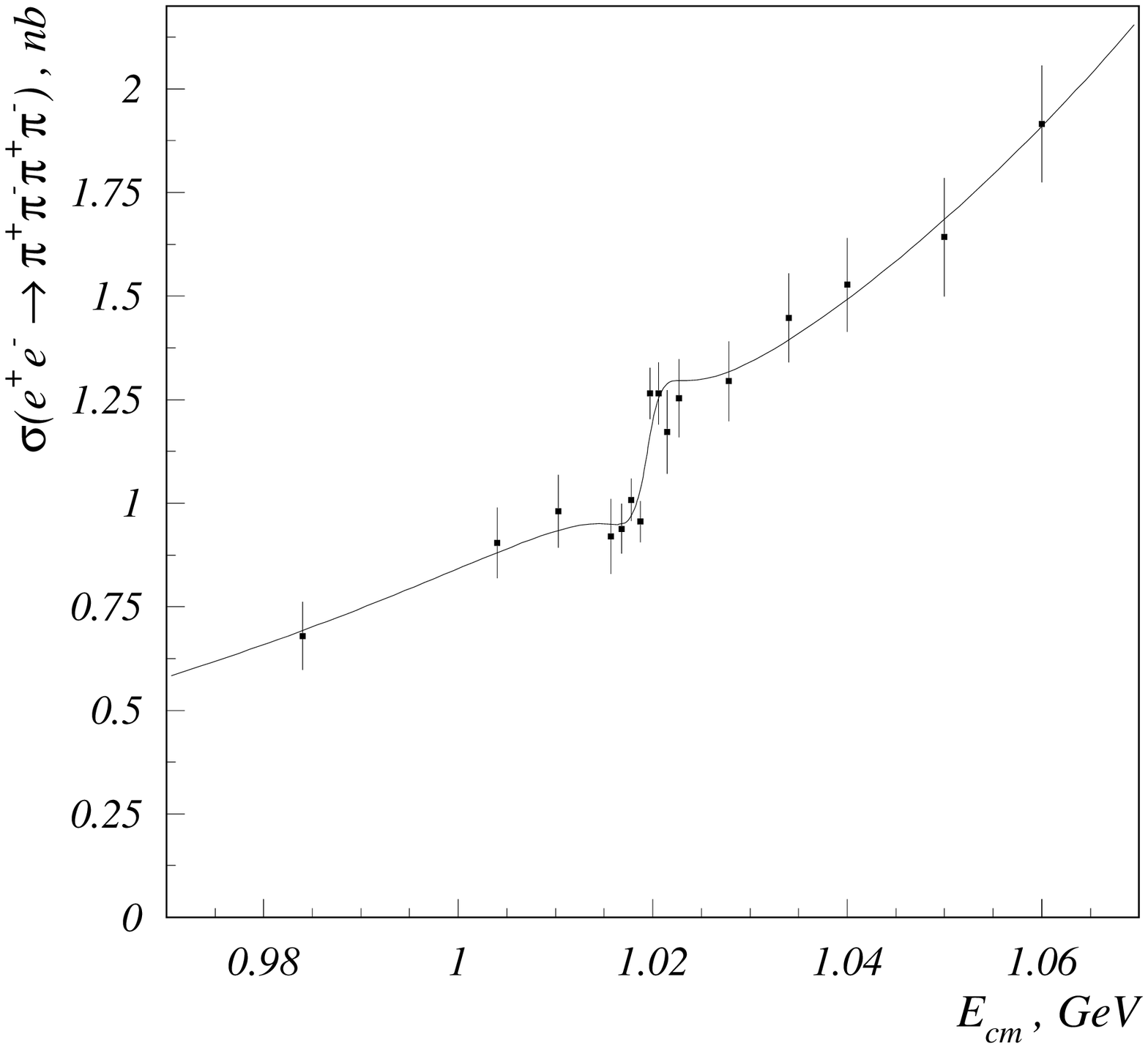}
     \caption{Cross section of the process \eefourpi\ in the $\phi$ meson
       region}
     \label{fig:xsphi}
\end{center}
\end{figure}

Table~\ref{tab:xsec} presents the summary of the
cross section calculations. Figure \ref{fig:xsphi} shows the energy
dependence of the cross section near the $\phi$ meson. 
The obtained values of the cross sections measured below and above
the $\phi$ meson match our previous results well \cite{our,rho4pi}.
Only statistical
errors are shown in Fig. \ref{fig:xsphi}.
The systematic uncertainty comes from the following sources:
\begin{itemize}
\item
selection criteria and background suppression - 11\%
\item
event reconstruction - 5\%
\item
detection efficiency dependence on the production mechanism - 3\%
\item
beam energy spread - 2\%
\item
radiative corrections - 1.6\%.
\item
luminosity determination - 1.5\%
\end{itemize}
The overall systematic uncertainty was estimated to be $\approx$ 13\%.

\begin{table}[ht]
\caption{Summary of the cross section calculations}
\label{tab:xsec}
\begin{center}
  \begin{tabular}{|c|c|c|c|}
    \hline
    $E_{cm}$, & $L$, & $N_{4\pi}$ & $\sigma$, \\
    GeV & $\mbox{nb}^{-1}$ & & nb \\
    \hline \hline
0.984   &  382.0  & 69   &  0.68 $\pm$ 0.08 \\
1.004   &  485.9  & 112  &  0.90 $\pm$ 0.09 \\
1.0103  &  503.4  & 125  &  0.98 $\pm$ 0.09 \\
1.0157  &  442.3  & 103  &  0.92 $\pm$ 0.09 \\
1.0168  &  1036.4 & 246  &  0.94 $\pm$ 0.06 \\
1.0178  &  1562.6 & 393  &  1.01 $\pm$ 0.05 \\
1.0187  &  1555.9 & 368  &  0.96 $\pm$ 0.05 \\
1.0197  &  1361.6 & 416  &  1.26 $\pm$ 0.06 \\
1.0206  &  923.7  & 280  &  1.27 $\pm$ 0.08 \\
1.0215  &  476.7  & 134  &  1.17 $\pm$ 0.10 \\
1.0227  &  584.6  & 177  &  1.25 $\pm$ 0.09 \\
1.0278  &  573.9  & 181  &  1.29 $\pm$ 0.10 \\
1.034   &  519.7  & 183  &  1.45 $\pm$ 0.11 \\
1.040   &  491.5  & 182  &  1.53 $\pm$ 0.11 \\
1.050   &  333.2  & 132  &  1.64 $\pm$ 0.14 \\
1.060   &  399.0  & 184  &  1.92 $\pm$ 0.14 \\
\hline
  \end{tabular}
\end{center}
\end{table}

The energy dependence of the cross section shown in Fig. \ref{fig:xsphi} 
demonstrates a clear interference pattern in the vicinity
of the $\phi$ meson. To describe the interference behavior we parameterize
the cross section according to the following formula:
\begin{eqnarray}
  \sigma_{\eefourpi}(E) & = & \sigma_0 f(E) \left| 1 - 
    Z \frac{m_{\phi}\Gamma_{\phi}}{m_{\phi}^{2}-E^2-iE\Gamma_{\phi}}
    \right|^2\,, \label{eq:fitex}
\end{eqnarray}
where $E=2E_{beam}$; $\sigma_0$ is the nonresonant cross section of the
process \eefourpi\ at $E=m_{\phi}$; $f(E)$ is a smooth function
describing the nonresonant behavior of the cross section
and normalized to 1 at $E=m_{\phi}$;
$m_{\phi}$ and $\Gamma_{\phi}$ are the $\phi$ meson mass and width; and $Z
= |Z| e^{i\psi}$ is a complex interference amplitude. The following
values of the parameters were obtained from the fit with $f(E) =
e^{A(E-m_{\phi})}$ ($A$ is a slope parameter):
\begin{eqnarray}
  \sigma_0 & = & 1.114 \pm 0.035 \pm 0.056 \quad \mbox{nb}, \nonumber \\
  \re Z & = & 0.122 \pm 0.027 \pm 0.033, \nonumber \\
  \im Z & = & -0.003 \pm 0.025 \pm 0.058 \quad . \label{eq:fitpar}
\end{eqnarray}
The obtained value of $\chi^2$/n.d.f. characterizing the fit quality was 
8.31/12  corresponding to a 75\% confidence level. 
Since the observed value of the resonance amplitude is about the
same order ($\sim 10\%$) as the statistical errors in 
any one of the measured
values of the cross section, the interference pattern in the cross
section behavior could appear due to statistical fluctuations
rather than the decay $\phi\to\fourpi$.
To check the consistency of the data with this assumption, we performed
a fit with the amplitude $|Z|$ fixed to 0. The obtained $\chi^2$/n.d.f. value
was 27.90/12  corresponding
to the 0.5\% confidence level. 
Thus, we can really claim evidence for the decay $\phi\to\fourpi$.

The first error in each fit parameter in (\ref{eq:fitpar}) is
statistical, while the second one is systematic.
Table \ref{tab:sys} lists the main sources of systematic uncertainties
contributing to the overall systematic error of the fit parameters.
Let us discuss some of these uncertainties in more detail.
\begin{table}[ht]
\caption{Systematic uncertainties of the fit parameters}
\label{tab:sys}
\begin{center}
  \begin{tabular}{|p{0.5\textwidth}|ccc|}
    \hline
    & \multicolumn{3}{c|}{Contribution to fit parameters} \\ \cline{2-4}
    Source of uncertainty &
    $\sigma_0$, nb & $\re Z$ & $\im Z$ \\
    \hline\hline
    Selection criteria  &
    0.019 & 0.013 & 0.046 \\
    \hline
    Radiative corrections &
    0.018 & 0.001 & $<$0.001 \\
    \hline
    Luminosity  and efficiencies &
    0.025 & 0.027 & 0.024 \\
    \hline
    Effective energies $\bar{E}$ &
    0.003 & 0.006 & 0.009 \\
    \hline
    Beam energy spread &
    0.001 & 0.011 & 0.003 \\
    \hline
    Shift of VEPP-2M energy scale &
    0.002 & 0.001 & 0.006 \\
    \hline
    Model dependence on $f(E)$ &
    0.042 & $<$0.001 & 0.024 \\
    \hline\hline
    Overall uncertainty &
    0.056 & 0.033 & 0.058 \\
    \hline
  \end{tabular}
\end{center}
\end{table}

The influence of the event selection procedure
on the fit parameters was estimated in the
following way. Three data samples were selected with an additional
requirement for background suppression and variations of the
criteria described in Section \ref{sec:analysis}:
\begin{enumerate}
  \item Application of a stricter requirement on the impact parameter of
    each track $r_{min} < 0.3$ cm 
    suppresses $\simeq$ 80\% of the background
    events from the reaction (\ref{eq:kskl}), leaving about 150 $K_S^0
    K_L^0$ events in the \fourpi\ data sample
  \item The probability $W_{\pi^+\pi^-}$  for two tracks
    with the smallest angle between them to be pions was calculated (see
    \cite{rho4pi} for more detail). The requirement $W_{\pi^+\pi^-} >
    0.5$ rejects $\approx$ 70\% of the background events from the
    processes (\ref{eq:3pi})--(\ref{eq:ppg})
  \item Restriction $\psi_{min} > 1.5$ suppresses the background from
    the reactions (\ref{eq:3pi})--(\ref{eq:ppg}) to the level of a few
    events.
\end{enumerate}
For all of these data samples the values of the cross section were
calculated at each energy point and the energy dependence of the
cross section was fit by the expression (\ref{eq:fitex}). The 
fit parameters thus obtained agree with the parameters
(\ref{eq:fitpar}) within the statistical errors.
Deviations from the values (\ref{eq:fitpar}) give an estimate for
the systematic uncertainty due to the event selection.
These estimates are shown in the first line of Table \ref{tab:sys}. 
The imaginary part of the interference amplitude $\im Z$ is very
sensitive to the resonant background. Thus, the presence of  events
of the reactions (\ref{eq:3pi}) and (\ref{eq:kskl}) in the final data
sample gives a major contribution (about 0.046) to
the systematic uncertainty of this parameter. The
calculation of radiative corrections \cite{KF} requires the value of
the threshold energy $\Delta E$ for the detection of radiated
photons. The uncertainty in $\Delta E$ gives about 1.6\%
contribution to the systematic error of $\sigma_0$.
Uncertainties in the luminosity determination, detection and
reconstruction efficiencies lead to  systematic uncertainties in
parameters $\sigma_0$, $\re Z$ and $\im Z$ comparable to statistical
errors.
The parameter $\sigma_0$ is sensitive to the choice of the function
$f(E)$, which approximates the nonresonant behavior of the cross
section in formula (\ref{eq:fitex}). A fit of the energy dependence of
the cross section was performed with the linear function $f(E)$
instead of an exponential  as in (\ref{eq:fitpar}). The systematic error
in $\sigma_0$ due to the choice of $f(E)$ was estimated to be about
3.8\%.

Using real and imaginary parts $\re Z$ and $\im Z$ of the 
interference amplitude $Z$ the values of $|Z|$ and $\psi$ were
calculated:
\begin{eqnarray}
  |Z| & = & 0.122 \pm 0.027 \pm 0.033, \nonumber \\
  \psi & = & ( -1 \pm 12 \pm 27)^{\circ}. \nonumber
\end{eqnarray}
The branching ratio of the decay $\phi \to \fourpi$ can be
calculated using the following expression:
\begin{displaymath}
  Br(\phi\to\fourpi) = \frac{\sigma_0 |Z|^2}{\sigma_{\phi}} = (3.93 \pm
  1.74 \pm 2.14) \cdot 10^{-6}\,,
\end{displaymath}
where $\sigma_{\phi} = 12\pi Br(\phi \to e^+e^-)/m_{\phi}^2 = 4224 \pm
113$ nb \cite{pdg} is the cross section of the $\phi$ meson production.
This is the first measurement of this quantity, and it supersedes the upper 
limit on the branching ratio of 8.7 $\cdot 10^{-4}$ obtained in Orsay 
\cite{dm1} as well as the upper limit of 1 $\cdot 10^{-4}$ placed by CMD-2 
and based on part of the whole data sample \cite{4picmd}.  

%===================================================================

\section{Search for decays $\phi\to\eta\pi^+\pi^-$ and \\
  $\phi\to\fourpi\pi^0$}

The same data sample of preselected events with four charged tracks
was used for the search of the decays $\phi\to\eta\pi^+\pi^-$,
$\eta\to\pi^+\pi^-\pi^0$ and $\phi\to\fourpi\pi^0$. 
In this analysis at least two photons detected in the
calorimeter were required.
The kinematic fit was performed taking into account energy ---
momentum conservation. In the reconstruction procedure
all charged particles  were assumed to be pions.

One of the main problems is additional (``fake'') photons induced by
the products of nuclear interactions of charged pions in the detector
material.
The following simple method was used to suppress  such fake
photons. 
In the kinematic fit the energy resolution of  
photons was loosened to the value $\sigma_{E_{\gamma}} = E_{\gamma}$ + 20 MeV. 
Thus, the photon energy was allowed to vary in a wide range during the fit. 
Only two photons were included in the fit. For events with more than two 
detected photons, the fit was repeated with all possible pairs of photons
and the pair with the smallest $\chi^{2}_{4\pi2\gamma}$ characterizing
the fit quality was selected.
Events with the reconstructed photon energy
below 30 MeV 
were rejected from the subsequent analysis.
The following requirements were additionally applied: $\chi^2_{4\pi2\gamma}$/
n.d.f. $<$ 10/4 and
the invariant mass of the photon pair is near the $\pi^0$ mass:
$|M_{\gamma\gamma}-m_{\pi^0}| < 30$ MeV.

The process (\ref{eq:kpkm}) in which products of kaon nuclear
interactions scatter back to the drift chamber and induce two extra tracks
or one of the kaons decays via the $K^{\pm}\to\pi^{\pm}\pi^+\pi^-$ channel,
accompanied by fake photons, can contribute to the background for the
decays under study. Another source of the background is the reaction 
\begin{eqnarray}
 e^+e^- \to \omega\pi^0\,,\omega \to \pi^+\pi^-\pi^0 
\label{eq:ompi0}
\end{eqnarray}
with the Dalitz decay of one of the neutral pions. 
The main background, however, comes from the process
(\ref{eq:kskl}) followed by the $K_S^0 \to \pi^+\pi^-$ and $K_L^0 \to
\pi^+\pi^-\pi^0$ decays. 

The same restrictions on the dE/dx of the tracks 
as those applied in the search for the decay $\phi \to \pi^+\pi^-\pi^+\pi^-$
(see Fig. \ref{fig:dedx}), were used to suppress the background from 
the decay (\ref{eq:kpkm}). 

To reject the background from the reaction (\ref{eq:ompi0}), we searched 
for a pair oppositely charged particles with the minimum space angle 
$\psi_{min}$ between the tracks. Assuming this pair to be $e^+e^-$ 
and taking the photon with the smaller energy, the invariant mass 
$M_{e^+e^-\gamma_{2}}$ was calculated. The requirements $\psi_{min} >$ 0.3 and 
$M_{e^+e^-\gamma_{2}} >$ 170 MeV reduced the background from the reaction
$e^+e^- \to \omega\pi^0$ to a negligible level:  $N_{\omega\pi^0} <$ 0.1.  

To reject the background from the decay (\ref{eq:kskl}) events in which
at least one of the $\pi^+\pi^-$ pairs satisfies the conditions
(\ref{eq:unselks}) were excluded. Additional suppression of
$K_S^0 K_L^0$ events was achieved by restricting the impact
parameter of each track: $r_{min} < 0.3$ cm.
Only $N^{vis}_{\eta\pi^+\pi^-} = 2$ candidate events survive after
applying these selection criteria.

Events for which the impact parameter of at least one track has the value 
$r_{min} >$ 0.3 cm are mostly coming from the decay (\ref{eq:kskl}).
The observed number of such events satisfying all above criteria
but $r_{min} <$ 0.3 cm is $N_{K^0_{S}K^0_{L}}$  = 6. 
Applying the whole set of selection criteria 
and requiring that for at least one $\pi^+\pi^-$ pair the conditions 
(\ref{eq:unselks}) are held,
one can obtain a practically pure $K^0_{S}K^0_{L}$ sample.
From the distribution of $r_{min}$ for thus selected events  
of the process (\ref{eq:kskl}) the ratio 
$N_{K^0_{S}K^0_{L}}(r_{min}<0.3)$/$N_{K^0_{S}K^0_{L}}(r_{min}>0.3)$=
0.20 $\pm$ 0.03 was obtained in good agreement with simulation. 
Using this ratio, the expected background in the region $r_{min}  <$ 0.3 cm
was estimated to be $N_{bg} = 1.2^{+0.7}_{-0.4}$.
Thus, the upper limit can be set on the number of signal events:
$N_{\eta\pi^+\pi^-} <$ 5.1 at 90\% CL \cite{feld}.    
The 90\% CL upper limit can be correspondingly obtained for 
the decay probability:
\begin{eqnarray}
  && Br(\phi\to\eta\pi^+\pi^-) 
   < \frac{N_{\eta\pi^+\pi^-}}
  {N_{\phi}Br(\eta\to\pi^+\pi^-\pi^0)\varepsilon_{\eta\pi^+\pi^-}} =
  1.8 \cdot 10^{-5},
\label{eq:etapipi}
\end{eqnarray}
where $N_{\phi} \approx 16 \cdot 10^6$ \cite{prep99} is the total
number of $\phi$ meson events recorded by CMD-2 in the experiment,
$Br(\eta\to\pi^+\pi^-\pi^0) = 0.231 \pm 0.005$ \cite{pdg} is the
branching ratio of the $\eta$ decay and $\varepsilon_{\eta\pi^+\pi^-} =
0.09 \pm 0.01$ is the detection efficiency obtained from simulation. 
To take into account the uncertainties in the $Br(\eta\to\pi^+\pi^-\pi^0)$
and $\varepsilon_{\eta\pi^+\pi^-}$, their values were lowered by one
standard deviation while calculating the upper limit in
(\ref{eq:etapipi}). 
This upper limit is approximately 15 times better than the previous one
also set by the CMD-2 group using the $\eta \to \gamma \gamma$
decay mode and based on part of the total data sample \cite{3pi}. 

Calculating the  detection efficiency for the process 
$e^+e^- \to \fourpi\pi^0$ under the assumption of the constant 
matrix element and skipping the probability of
the decay $\eta \to \pi^+\pi^-\pi^0$ in (\ref{eq:etapipi}), one can
obtain the upper limit on the branching ratio of the direct decay
$\phi\to \fourpi\pi^0$:
\begin{displaymath}
  Br(\phi\to\fourpi\pi^0) < 4.6 \cdot 10^{-6}
  \quad 90 \% \,\mbox{CL}. 
\end{displaymath}
This limit is 25 times better than the previous one
placed in Ref. \cite{cmd}.

%===================================================================

\section{Discussion}

At the present time no theoretical calculations exist for the value 
$Br(\phi\to\fourpi)$.
A simple estimate can be performed \cite{ach-pc} taking into account 
the $\phi-\gamma$ transition:
\begin{eqnarray}
&& Br(\phi\to\gamma^\ast\to\fourpi) = 9\cdot\frac{(Br(\phi\to
  e^+e^-))^2}{\alpha^2}\cdot\frac{\sigma_0} {\sigma_{\phi}} = 3.99 \cdot
  10^{-6} \nonumber \,,
\end{eqnarray}
where the values $\sigma_0 = 1.114$ nb 
and $\sigma_{\phi} = 4224$ nb were used for
the nonresonant cross section of the process \eefourpi\ and 
the total cross section of the $\phi$ meson production respectively.
The measured branching ratio is consistent
with this estimate. 

Note that this measurement is also of interest to clarify
the problem of two conflicting results for the
branching ratio of the related decay $\phi \to \pi^+\pi^-$ recently 
measured by CMD-2 \cite{pipicmd} and SND groups \cite {pipisnd}.
While in the CMD-2 measurement the imaginary part
of the interference amplitude is consistent with zero, SND
claims a statistically significant non-zero imaginary part for this 
quantity.
The result obtained in our work for the $\phi \to \pi^+\pi^-\pi^+\pi^-$  
decay mode does not contradict a purely one-photon mechanism of the
$\rho-\phi$ mixing. However, large systematic uncertainties preclude
unambiguous conclusions.

In papers \cite{ach-karn,ach-kozh} 
using various models of $\phi$--$\rho$
and $\phi$--$\omega$ mixing,
the value for the branching
ratio of the decay $\phi\to\eta\pi^+\pi^-$ was calculated: 
\begin{displaymath}
  Br(\phi\to\eta\pi^+\pi^-) = 0.35 \cdot 10^{-6}.
\end{displaymath}
This value is 50 times lower than the obtained upper
limit.

%===================================================================

\section{Conclusion}

The reaction \eefourpi\ has been studied in the energy range 0.984 to 1.06
GeV. About 3300 \fourpi\ events were detected. For the first time the
interference behavior of the cross section has been observed in the
vicinity of the $\phi$ meson. The branching ratio of the decay
$\phi\to\fourpi$ suppressed by $G$-parity conservation and OZI rule
has been measured:
\begin{displaymath}
  Br(\phi\to\fourpi) = (3.93 \pm 1.74 \pm 2.14) \cdot 10^{-6}.
\end{displaymath}
Upper limits have been set on the branching ratios of the decays 
$\phi\to\fourpi\pi^0$ and $\phi\to\eta\pi^+\pi^-$:
\begin{eqnarray}
  Br(\phi\to\fourpi\pi^0) & < & 4.6 \cdot 10^{-6}\quad 90\%\,\mbox{CL},
  \nonumber \\
  Br(\phi\to\eta\pi^+\pi^-) & < & 1.8 \cdot 10^{-5}\quad 90\%\,\mbox{CL}.
  \nonumber
\end{eqnarray}

The authors are grateful to the staff of VEPP-2M for excellent
performance of the collider, to all engineers and technicians who
participated in the design, commissioning and operation of CMD-2.
Special thanks are due to N.N.Achasov for useful discussions.

%===================================================================

\end{document}